\documentclass [twocolumn,epsfig,prl,10pt,floatfix,amsmath,amssym,showpacs]{revtex4}
\usepackage[dvips]{epsfig}

\usepackage{float}

\begin{document}

\title{Chirality and Equilibrium Biopolymer Bundles}
\author{Gregory M. Grason}
\author{Robijn F. Bruinsma}
\affiliation{Department of Physics and Astronomy, University of California at Los Angeles, Los Angeles, CA 90024, USA}

\begin{abstract}
We use continuum theory to show that chirality is a key thermodynamic control parameter for the aggregation of biopolymers: chirality produces a stable disperse phase of hexagonal bundles under moderately poor solvent conditions, as has been observed in {\it in-vitro} studies of F-actin [O. Pelletier {\it et al.}, Phys. Rev. Lett. {\bf 91}, 148102 (2003)]. The large characteristic radius of these chiral bundles is not determined by a mysterious long-range molecular interaction but by in-plane shear elastic stresses generated by the interplay between a chiral torque and an unusual, but universal, non-linear gauge term in the strain tensor of ordered chains that is imposed by rotational invariance.
\end{abstract}
\pacs{87.16.Ka, 64.75.+g, 61.30.Dk}
\date{\today}

\maketitle

The structural integrity of cells depends on long, semi-flexible biopolymers--such as filamentous actin (F-actin), microtubules, and intermediate filaments--that make up the cytoskeleton. The controlled assembly and disassembly of these polymers is central to the division, development, and locomotion of cells~\cite{alberts}.  {\it In-vitro} studies of their thermodynamic, structural, and elastic properties have contributed importantly to our understanding of the physical mechanisms that underlie the functioning of the cytoskeleton. For example, F-actin filaments adopt characteristic structural motifs inside cells: open, branched networks and filament bundles \cite{theriot}. Growing networks play an important role during cell spreading while dense bundles are a necessary component of cellular force generation (Òstress fibersÓ) and cell protrusions (e.g. microvilli and stereocilia). Networks and bundles appear as separate regions in the phase diagram of solutions of actin proteins and linkers under {\it in-vitro} conditions,~\cite{tang_pchem_96, safinya_prl_03, pelletier_prl_03,  angelini_pnas_03} and much is now understood about the function of specialized linker proteins in the structural organization of actin filaments.  Interestingly, even non-structural biopolymers like DNA~\cite{bloomfield} can exhibit these morphologies in solutions with generic linker units, such as polyvalent ions.

The observation of stable biopolymer bundle aggregates poses an interesting puzzle.  Equilibrium solutions of (charged) rods of fixed length are predicted to form networks at low linker concentrations ~\cite{borukov_pnas_05}. Beyond a critical linker concentration, cylindrical aggregates should form without a limit on their diameter:  bundles can grow in size until the supply of free rods is depleted. Explanations for the observed finite size of the bundles have focused on the possibility of long-range interactions~\cite{henle_pre_05}, on kinetic limitations to the bundle size~\cite{ha_liu_epl_99}, and on assembly defects of toroidal-shaped biopolymer aggregates~\cite{gelbart_biophys_98}. A recent study~\cite{wong_prl_07} of fluorescently labeled F-actin mixtures has cast doubt on these explanations: non-toroidal bundles were found to maintain a well-defined, fixed diameter--in the range of 40-100 nm--that was greatly in excess of the electrostatic screening length even when filament exchange between and along different bundles was rapid. 

\begin{figure}[b]
\center \epsfig{file=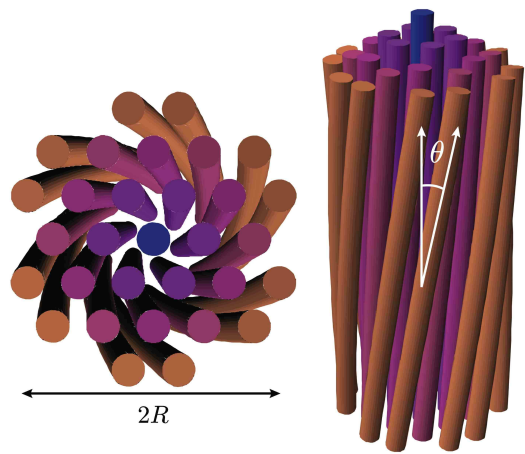, width=2.75in}\caption{Top view (left) and side view (right) of twisted bundle of hexagonally-ordered filaments.  For a torsional strain, $\Omega$ , the outer most filaments make an angle $\theta=\Omega R$ with the central axis.  Hence, elastic strainÑchain bending and lattice shearÑbuild up as twisted bundle grows.} 
\label{fig: twisted}
\end{figure}
 
It is the aim of this Letter to argue that {\it chirality} is a key thermodynamic control parameter for biopolymer bundle aggregation. Chirality already is well known to play a determining role for the phase diagram of {\it soluble} chiral biopolymers. It is, for example, responsible for both the cholesteric liquid crystal and the ``blue" phases of concentrated DNA solutions~\cite{livolant}.   Here, we show that chirality also plays a key role for biopolymers in Òpoor solventÓ conditions: in the presence of chirality, a stable {\it disperse} phase of twisted, finite-sized bundles appears, separating the bulk aggregation phase from the single filament (or network) phase. The large characteristic dimensions of these chiral bundles is {\it not} determined by a mysterious long-range molecular interaction but, instead, by in-plane shear elastic stresses generated by the competition between a chiral torque and a non-linear gauge term appearing in the strain tensor for ordered chains.

To demonstrate these claims, we will not construct an explicit model of F-actin based on specific assumptions about molecular structure, instead, we will rely on symmetry arguments to construct the generic free energy of a hexagonal bundle of chiral, semi-flexible polymers of fixed length, $L$, where chains can slide freely along the local hexagonal axis.  Deformations of the array are described by the in-plane chain displacement field  ${\bf u}_\perp({\bf r}_\perp,z)$ of chains located at  ${\bf r}_\perp$ (in the plain) as a function of height, $z$.  Local chain-tangent unit vectors are given by ${\bf t} \simeq 1+ \partial_z {\bf u}_\perp$ and local curvature of the chains is given by  $\kappa=|\partial^2_z {\bf u}_\perp|$. The rotationally-invarient in-plane strain tensor of the array is determined by ${\bf u}_\perp$ through
\begin{equation}
\label{eq: uij}
u_{ij}=\frac{1}{2} \big(\partial_i u_{\perp j}+\partial_j u_{\perp i}-\partial_i {\bf u}_\perp \cdot \partial_j {\bf u}_\perp - t_i t_j\big) \ ,
\end{equation}
where indices refer only to in-plane directions.  The first three terms are familiar from the classical theory of the elasticity of two-dimensional solids.  The last term is a {\it gauge term} imposed by the chain connectivity: a uniform rotation of the chains away from the bundle axis reduces their mutual perpendicular separation, generating strain~\cite{selinger_bruinsma_pra_91}.  The elastic strains captured by the gauge contribution to $u_{ij}$ play a special role in the following.

The deformation energy is the sum of an in-plane energy ${\cal H}_\perp$  that depends only on this strain tensor.  For a chiral bundle with hexagonal symmetry, ${\cal H}_\perp$ can be expanded to second order in $u_{ij}$ as:
\begin{equation}
\label{eq: Hperp}
{\cal H}_\perp=\frac{1}{2} \int d^3 x \Big\{\lambda_\perp u^2_{ii} +2 \mu_\perp u_{ij} u_{ij} \Big\} \ ,
\end{equation}
plus an out-of-plane term ${\cal H}_{\bf t}$ that can be expanded to second order in ${\bf t}$ as~\cite{note1}:
\begin{multline}
\label{eq: Ht}
{\cal H}_{\bf t}=\frac{1}{2} \int d^3 x \Big\{K_1(\nabla_\perp \cdot {\bf t})^2 \\ +K_2(\nabla_\perp \times {\bf t}-\hat{z}/P)^2 +K_3(\partial_z {\bf t})^2 \Big\} \ .
\end{multline}
The in-plane and out-of-plane contributions to the deformation energy are linked, both by the relation  ${\bf t}\simeq\hat{z}+\partial_z {\bf u}_\perp$ between the tangent vector and the in-plane displacement, and by the non-linear gauge term that appears in the strain tensor eq. (\ref{eq: uij}).  In the conclusion we will discuss estimates for the Lam\'e constants $\lambda_\perp$  and $\mu_\perp$ and the Frank constants $K_{1,2,3}$.  In what follows we assume, $K_3 \gg K_{1,2}$ .  Below we focus on bundles of variable radius $R$ and fixed length $L$, such that $L\gg R$.

The term $K_2 P^{-1} (\nabla_\perp \times {\bf t})_z$ has the form of an {\it external torque} that tends to twist a bundle. We will show elsewhere that the resulting in-plane displacement field can be represented, to a very good approximation, by a pure torsional deformation where each cross-section of the bundle is rotated by rigid-body rotation over a certain angle with respect to the sections above and below it (see Fig. \ref{fig: twisted}).  Near the cross-section at $z=0$ , such a torsional displacement field can be written as:
\begin{equation}
\label{eq: utwist}
{\bf u}_\perp (r, z) \simeq r (z \Omega) \hat{\phi} - \frac{r(z \Omega)^2}{2} \hat{r} \ , 
\end{equation}
to second order in the {\it torsional strain}, $\Omega$ .  The deformation energy per unit volume  $E_{elast}(R,\Omega)$ for given $\Omega$  and bundle radius $R$  can be obtained by inserting eq. (\ref{eq: utwist}) into ${\cal H}_\perp$ and ${\cal H}_{\bf t}$ :
\begin{equation}
\label{eq: E1}
E_{elast}(R,\Omega)=\bar{\mu}_\perp (\Omega R)^4+\bar{K}_2 \Omega^2+\bar{K}_3 \Omega^4 R^2-\gamma \Omega \ .
\end{equation}
Here,  $\bar{\mu}_\perp=25 \mu_\perp/216$, $\bar{K}_2=2K_2$, $\bar{K}_3$ and $\gamma = 2K_2/P$ (we have assumed the array to be incompressible). We assume the magnitude $\theta=\Omega R$  of the strain to be small compared to one, as we are neglecting higher-order terms in ${\cal H}_\perp$. The last three terms of eq. (\ref{eq: E1}) result from a straightforward evaluation of  (\ref{eq: Ht})  while the first term, which is non-linear and dominates for large $R$, is the shear stress produced by the gauge term in eq. (\ref{eq: Hperp}). Figure \ref{fig: twisted} shows the physical origin of this term.  Chains far from the central axis are increasingly tilted away from the bundle axis by the torsional strain field, leading to compression of the local hexagonal packing of chains along direction, $\hat{r} \times {\bf t}$.  The term, $-\gamma \Omega$, in eq. (\ref{eq: E1}) acts as a {\it chiral} contribution to the aggregation energy density of the rod, with the first three terms the unavoidable energy penalty associated with chiral aggregation. There is, of course, a conventional achiral contribution as well to the aggregation energy density, which we denote as $-\epsilon$.
 
For  $K_2\ll K_3$ , the elastic energy is minimized by a torsional strain that depends on the bundle size as  $\Omega_0(R)=(\gamma/4 \bar{\mu}_\perp)^{1/3}R^{-2/3}(R^2+\lambda^2_{3\perp})^{-1/3}$, where $\lambda_{3\perp}=\sqrt{\bar{K}_3/\bar{\mu}_\perp}$  is an important characteristic length.  The total free energy of the bundle is the sum of the elastic energy eq. (\ref{eq: E1})--with $\Omega=\Omega_0(R)$ --the achiral aggregation energy and a surface energy associated with the reduced binding energy of filaments located on the surface of the bundle exposed to solution:
\begin{equation}
\label{eq: DF}
\frac{\Delta F(R)}{L}=-f_0 \frac{R^{4/3}}{(\lambda^2_{3\perp}+R^2)^{1/3}}-\pi R^2 \epsilon+2 \pi R \Sigma \ .
\end{equation}
Here, $f_0=3 \pi (\gamma^4/16 \bar{\mu}_\perp)^{1/3}$ and $\Sigma$ is the surface energy per unit area.

\begin{figure}
\center \epsfig{file=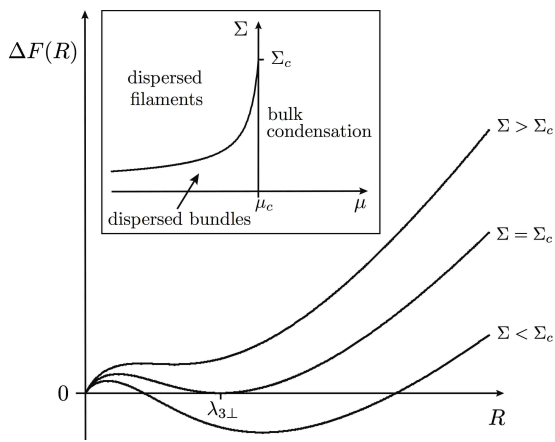, width=2.85in}\caption{The dependence of the free-energy of a chiral bundle on bundle radius, $R$, for the columnar phase of aggregates at the bulk condensation point ($\mu=\mu_c$).  When surface tension is greater than $\Sigma_c$  the free-energy minimum occurs for $R=0$ (single filaments).  For $\Sigma\leq\Sigma_c$ the minimum occurs at a finite radius, $R_0\geq \lambda_{3\perp}$ .  The inset shows the phase diagram for the columnar phase of aggregates as function of filament chemical potential and surface tension.} 
\label{fig: freeenergy}
\end{figure}

Under conditions of thermodynamic equilibrium, the fraction $P(R)$  of chains that will aggregate into bundles of radius $R$--the cluster size distribution--in chemical equilibrium with the surrounding solution follows the Boltzmann distribution~\cite{safran}:
\begin{equation}
\label{eq: P(R)}
P(R) \propto R^2 \exp \bigg\{-\frac{\Delta F(R)-\mu \rho_0 \pi R^2}{k_B T} \bigg\} \ .
\end{equation}
Here, $\mu$ the chemical potential of filaments and $\rho_0$ the area density of chains.  In the $L\rightarrow \infty$  limit, this cluster size distribution is dominated by the minima of $\Delta F(R)$. Near the point of bulk condensation--when $\mu \lesssim \mu_c \equiv -\epsilon L/ \rho_0$--and for large surface energies, the only minimum is an $R=0$ boundary minimum for which $\Delta F=0$, corresponding to individual chains in solution or small clusters with an exponential size distribution. As the surface energy is reduced, a second minimum develops at $R=R_0(\Sigma)$ with a width $\delta R \propto L^{-1/2}$.  For large  $L$, an appreciable fraction of the chains can only condense into the clusters associated with this minimum provided that  $\Delta F(R_0) -\mu \rho_0 \pi R_0^2 <0$, otherwise the boundary minimum at $R=0$ dominates the distribution.  Hence, a thermodynamically stable disperse phase occurs when the surface energy drops below a critical value $\Sigma_c$. When $\mu=\mu_c$, $\Sigma_c$ is given by:
\begin{equation}
\label{eq: sigmac}
\Sigma_c=\frac{f_0}{2^{4/3} \pi \lambda^{1/3}_{3\perp}} \ ; \ \ R_0(\Sigma_c)=\lambda_{3 \perp} \ .
\end{equation}
For $\Sigma < \Sigma_c$ , $\Delta F(R_0)$ is negative and the second minimum entirely dominates the cluster distribution function in the large $L$  limit, and this corresponds to the stable dispersed phase of bundles (see Fig. \ref{fig: freeenergy}). The peak position shifts to values larger than $\lambda_{3 \perp}$ as the surface energy is decreased, and in the limit of low surface energies it grows as  $R_0 \sim \gamma^4/\bar{\mu}_\perp \Sigma^3$. Thus, $\lambda_{3 \perp}$ corresponds to a minimum bundle size at the bulk condensation point. Note that this minimum bundle size is determined by the non-linear contribution to the elastic energy, and it diverges in the $\bar{\mu}_\perp \rightarrow 0$  limit.  Away from the bulk condensation point, for  $\mu<\mu_c$, the minimum in $\Delta F(R) -\mu \rho_0 \pi R^2$  is reduced by the bulk free-energy penalty, leading to reduction of the critical surface tension.  Sufficiently far from the bulk condensation point we find a critical surface tension, $\Sigma_c \simeq f_0^{3/2}3^{-1/2}\lambda^{-1}_{3 \perp} (\mu_c-\mu)^{-3/2}$.  The phase behavior of columnar aggregates is shown in Figure \ref{fig: freeenergy} (inset).

If the aggregate adopts full three-dimensional positional order, then chains cannot slide past each other and a new term determined by the out-of-plane shear modulus $\mu_{\parallel}$  must be added deformation energy density:
\begin{equation}
\label{eq: E2}
E_{elast}(R,\Omega)=\bar{\mu}_\parallel (\Omega R)^2+\bar{\mu}_\perp (\Omega R)^4+\bar{K}_3 \Omega^4 R^2-\gamma \Omega \ ,
\end{equation}
where $\bar{\mu}_\parallel=\mu_\parallel/4$ and we have here dropped the term proportional to $\bar{K}_2$.  This new term is actually familiar from the classical theory of elasticity.  It is the strain energy density of an elastic rod under uniform torsion ~\cite{landau_lifshitz}.  In the physical range, where the magnitude of the elastic strain $\theta = \Omega R$ must be less than one, the new harmonic term $\bar{\mu}_\parallel \theta^2$  dominates--for $\bar{\mu}_\parallel\approx \bar{\mu}_\perp$--over the anharmonic term $\bar{\mu}_\perp \theta^4$ that previously played such a central role. The form of  $P(R)$ for a bundle in the solid phase is somewhat more complex, but the key features can be illustrated by analyzing the limit of large $R$. 

If we minimize only the harmonic (first) and chiral (last) terms of eq. (\ref{eq: E2}) with respect to $\Omega$,  then the optimal strain and the minimum energy density depend on $R$  as $\theta_0(R) \simeq \gamma/ 2 \bar{\mu}_\parallel R$ and $E_{elast} \simeq -\gamma^2/  2 \bar{\mu}_\parallel R^2$, respectively, where we must demand $R\gg \gamma / \bar{\mu}_\parallel$ in order for $\theta \ll 1$.   The two anharmonic terms, $\bar{K}_3 \theta^2 /R^2$ and $\bar{\mu}_\perp \theta^4$, are then both small compared to $ -\gamma^2/  2 \bar{\mu}_\parallel R^2$ for large $R$.  However, in this regime the lowest order term contributes only the negative constant, $ -\pi \gamma^2/  2 \bar{\mu}_\parallel$, to  $\Delta F(R)$.  While this constant is important in terms of competition of chiral aggregates with individual filaments, it alone leads to no preferred bundle size.  Thus, to harmonic order, elasticity theory of three-dimensional chiral polymer solids predicts no maximum in $P(R)$ .  However, if we include anharmonic corrections, for sufficiently low surface energies and near to bulk condensation ($\mu \gtrsim \mu_c$), then there is a negative minimum  $\Delta F(R)$, and thus a dominant maximum  in $P(R)$ near  $R_0\sim (\bar{K}_3/ \Sigma)^{1/5} (\gamma/\bar{\mu}_\parallel)^{4/5} $  when $R_0 \ll \lambda_{3 \perp}$  and $R_0\sim (\bar{\mu}_\perp / \Sigma)^{1/3} (\gamma/\bar{\mu}_\parallel)^{4/3}$  for $R_0 \gg \lambda_{3 \perp}$ .  Importantly, in both hexagonal-columnar and hexagonal-solid phases, preferred bundle sizes are {\it only} possible if the effects of anharmonic elasticity are taken into account.

To be clear, our approach is generic, relying on symmetry arguments to dictate the form of the elastic energy of ordered chains, rather than a detailed, microscopic description of intermolecular forces between biopolymers.    Nevertheless, in order to apply the continuum theory to F-actin bundle formation, we need estimates for the elastic moduli and the Frank constants for realistic biopolymer bundles in solution.  For a bundle held together by linker units--proteins, cations, {\it etc.}--the Lam\'e constants $\lambda_\perp$ and $\mu_\perp$ can be crudely estimated to be of order $\Delta E/\ell d^2$ , with $\Delta E \sim 1-10 k_B T$  the binding energy per linker unit, $\ell$  the spacing between linkers along the polymer backbone, and $d \sim 1-10$ nm a typical molecular length (e.g., the spacing between the chains). The respective Frank constants $K_1$  and $K_2$  for splay and twist will be estimated as $\Delta E/\ell$ .  The Frank constant for {\it bend} is determined not primarily by linkers, but by the bending stiffness of the individual chains through through $K_3\sim k_B T \ell_p/d^2$  with $\ell_p$  the persistence length.  For F-actin $\ell_p$  is of the order of 10 $\rm{\mu}$m so indeed $K_3 \gg K_{1,2}$ for F-actin filaments as claimed earlier.  Given these estimates, the minimum size of bundles in the hexagonal columnar phase at the onset of aggregation,  $\lambda_{3 \perp} \sim \sqrt{d \ell_p} \sim$ 100 nm, indeed is of the order of the observed bundle size \cite{wong_prl_07}.  The surface energy will be estimated by  $\Sigma\sim \Delta E^*/\ell d$, where, in general,  $\Delta E^*$ can be less than $\Delta E$, for example, when surfactants are absorbed on the bundle surface.  The ratio $\Sigma_c(\mu_c)/\Sigma$ is of order  $(d/P)^{4/3}(d/\lambda_{3 \perp})^{1/3}\Delta E/\Delta E^*$.  When the disperse columnar phase is stable, this parameter must be be larger than one.  Disperse bundles are thus stabilized by reduced persistence lengths, by the action of surfactants, and by a short preferred chiral pitch $P$, preferably on the order of molecular scales.  For large persistence lengths, the radius of disperse bundles with solid order scales as $R_0 \sim (\bar{K}_3/ \Sigma)^{1/5} (\gamma / \bar{\mu}_\parallel)^{4/5}  \sim \ell_p^{1/5}$, which is considerably reduced in comparison to the scaling of minimum size for columnar bundles, $\lambda_{3\perp} \sim \ell_p^{1/2}$.  This indicates the radii of solid bundles is much less than that of the columnar phase, owing to the increased resistance to torsion~\cite{note}.

Confirmation of the description presented in this paper demands the observation of torsion in bundles of chiral biopolymers by microscopy. Cryo-AFM images of F-actin filaments by Shao {\it et al}. provided the first clear evidence of twisting~\cite{shao_biophys_00}. More recently, Ikawa {\it et al.} carried out a careful AFM study of F-actin filaments for different Mg$^{2+}$ ion concentrations ~\cite{ikawa_prl_07}.  For low Mg$^{2+}$ ion concentration, a network is observed which breaks up into braided bundles of a small number of filaments at increased linker concentrations.   Finally, synchrotron X-ray studies of F-actin bundles show small but clear deviations from perfect hexagonal packing~\cite{tang_pchem_96, angelini_pnas_03}, which would be consistent with the anharmonic chiral strain discussed in this paper. The combination of these observations and the theoretical analysis of the present paper we believe provides mounting evidence for chirality as a determining factor for the formation of thermodynamically stable F-actin bundles of finite size.

We conclude by noting that the present analysis did not allow for the possibility of {\it topological defects}. Kamien and Nelson~\cite{kamien_nelson}  (KN) showed that an infinite hexagonal columnar array of strongly chiral polymers is unstable against break-up into finite-sized regions of localized double-twist.  In this so-called ``moir\'e phase", double-twisting bundles are arranged in a triangular lattice, with neighboring bundles meeting along a tilt-grain boundary, necessary to accommodate the mismatch in chain orientation between adjacent bundles. The size scale $\xi_m$  of the bundle is determined by the repulsive energy between screw dislocations along the grain boundary and the intrinsic torsional stress as discussed in the present paper.  We speculate that the KN moir\'e phase could be viewed as the result of bundle-bundle aggregation produced at low surface energy.  As $\Sigma$ is decreased, and the equilibrium bundle size $R_0$  approaches $\xi_m$  from below, individual merge, ÒhidingÓ surface exposed to unfavorable solvent contact by introducing localized grain boundaries where neighboring bundles meet.  Alternatively, we may view the moir\'e phase as a precursor to the chirality-induced break up of the hexagonal columnar phase. In a future study, we aim to extend our analysis to the more general case including screw dislocations as well as to the case of the toroidal geometry observed for bundles of long DNA.

\begin{acknowledgments}
It is a pleasure to acknowledge many useful discussions with R. Kamien, S. Safinya and M. Henle.  This work was supported by the NSF under DMR Grant 04-04507.
\end{acknowledgments}


\begin{thebibliography}{99}
\bibitem{note}
The stability of the dispersed phase requires that $\gamma \theta > 2 \Sigma$ from which we have that $\Sigma/\gamma \ll 1$.  The scaling analysis of the solid phase shows that the mean size of dispersed bundles is smaller than columnar aggregates by a factor least as small as $ (\Sigma/\gamma)^{8/3} \ll 1$.
\end{thebibliography}
\end{document}